\documentclass[10pt,conference]{IEEEtran}

\IEEEoverridecommandlockouts
    \usepackage[pdftex]{graphicx}
    \graphicspath{{../pdf/}{../jpeg/}}
    \DeclareGraphicsExtensions{.pdf,.jpeg,.png}
    
    \usepackage[cmex10]{amsmath}
    \usepackage{hyperref}
    \usepackage{algorithm} 
    \usepackage{algpseudocode}
    \usepackage{array}
    \usepackage{mdwmath}
    \usepackage{mdwtab}
    \usepackage{eqparbox}
    \usepackage{comment}
    \usepackage{subfigure}
    \usepackage{romannum}
    \usepackage{graphicx} 
    \usepackage{booktabs} 
    \usepackage{cite}
    \usepackage{caption, multirow, makecell}
    \usepackage{siunitx}
    \usepackage{mhchem}
    \usepackage{float}
    \usepackage{adjustbox}
    \usepackage{multirow}
    \usepackage{amssymb}

\usepackage[T1]{fontenc}

\usepackage{tabularx} 
\IEEEoverridecommandlockouts

\begin{document}

\title{\textit{ProtoAoA}: Few-Shot Angle-of-Arrival Estimation using Prototypical Networks}

\author{\IEEEauthorblockN{ 
Elsayed Mohammed\IEEEauthorrefmark{3},
Omar Mashaal\IEEEauthorrefmark{3},
Alec Digby\IEEEauthorrefmark{1}, Pasquale Leone\IEEEauthorrefmark{1},
Lorne Swersky\IEEEauthorrefmark{1},\\
Ashkan Eshaghbeigi\IEEEauthorrefmark{1}, and
Hatem Abou-Zeid\IEEEauthorrefmark{3}
}
\IEEEauthorblockA{\IEEEauthorrefmark{3}{Department of Electrical and Software Engineering}, 
{University of Calgary}, Canada}
\IEEEauthorblockA{\IEEEauthorrefmark{1}{Qoherent Inc.}, 
{Toronto, Ontario, Canada}} 

\thanks{
This research was funded by Alberta Innovates and the Natural Sciences and Engineering Research Council of Canada (NSERC) through the NSERC Alliance-Alberta Innovates Program and the NSERC Discovery Grant RGPIN-2021-04050.}

}

\maketitle

\begin{abstract}

Angle-of-arrival (AoA) estimation is a crucial function in wireless communications used for localization, beam-forming, interference management, and other applications.  
Deep learning (DL) solutions have been proposed for AoA to mitigate limitations of traditional AoA estimation techniques such as sensitivity to noise and the inability to generalize across different array characteristics. A challenge, however, of DL-based approaches is their reliance on large data collection campaigns and model training. This paper proposes the application of Prototypical Networks (PN) to address this challenge and utilizes a real-world dataset collected on a software defined radio (SDR) testbed to validate the effectiveness of the proposed solution.
Prototypical Networks excel in extracting representative embeddings from unstructured input data, establishing class prototypes during training that can be few-shot trained on unseen classes. We demonstrate the efficacy of PNs for AoA classification using complex IQ samples, focusing on its ability to  correctly classify new, unseen angles that the model was not trained on previously. Our results show that training our proposed ProtoAoA on only 23\% of the AoA dataset classes can attain a mean absolute error (MAE) of 3$^{\circ}$ with only 4-shots of training on the unseen angles $-$ and an MAE of 2$^{\circ}$ with 32-shots of training data.
These results demonstrate that the developed prototypical network architecture requires remarkably few data samples to achieve reliable AoA estimation $-$ and highlights its potential for other wireless applications where data availability is limited. 

\end{abstract}

\IEEEoverridecommandlockouts

\begin{IEEEkeywords}
Angle of arrival, prototypical networks, few-shot learning, sparse data learning.
\end{IEEEkeywords}

\IEEEpeerreviewmaketitle


\section{Introduction}

Angle-of-arrival (AoA) estimation is a signal processing technique used in wireless communications to determine the direction from which a signal reaches an antenna array. This technique helps identify the precise location or direction of signal source. Therefore, it is crucial for applications such as beam-forming, localization, radar systems, and interference management. In cellular systems \cite{AoA-cell}, AoA estimation plays a significant role in enhancing signal quality, and improving overall communication efficiency. As a result, AoA estimation remains a key area of research, with focus in developing accurate, adaptable, and efficient methods suitable for various dynamic environments. More broadly, recent work has highlighted that hardware heterogeneity, codebook mismatch, and environment shift remain major obstacles to the practical generalization of ML-based wireless learning methods, including beam management systems \cite{BeamHeterogeneity}. Related efforts such as ProtoBeam further illustrate the need for adaptable wireless learning frameworks that can generalize across unseen antenna configurations \cite{ProtoBeam}.

Due to their ability to extract meaningful features from unstructured data, data-driven machine learning (ML) solutions have been proposed in different communication networks and signal processing tasks including AoA estimation.
The works in \cite{ML-AoA, AoA-cell, AoA-net} provide various data and artificial intelligence (AI) driven frameworks for AoA estimation. 
Such AI models can achieve super-resolution with lower complexity and signal snapshots that traditional AoA estimation techniques, and can be resilient to antenna perturbations, phase and amplitude inconsistencies, and can be antenna-agnostic. However, a key challenge in building these AI-driven solutions is the need for comprehensive large data collection and model training campaigns to achieve this robust performance.

\begin{figure}[t!] 
\centering

\includegraphics[width=3.6in, height=2.25in]{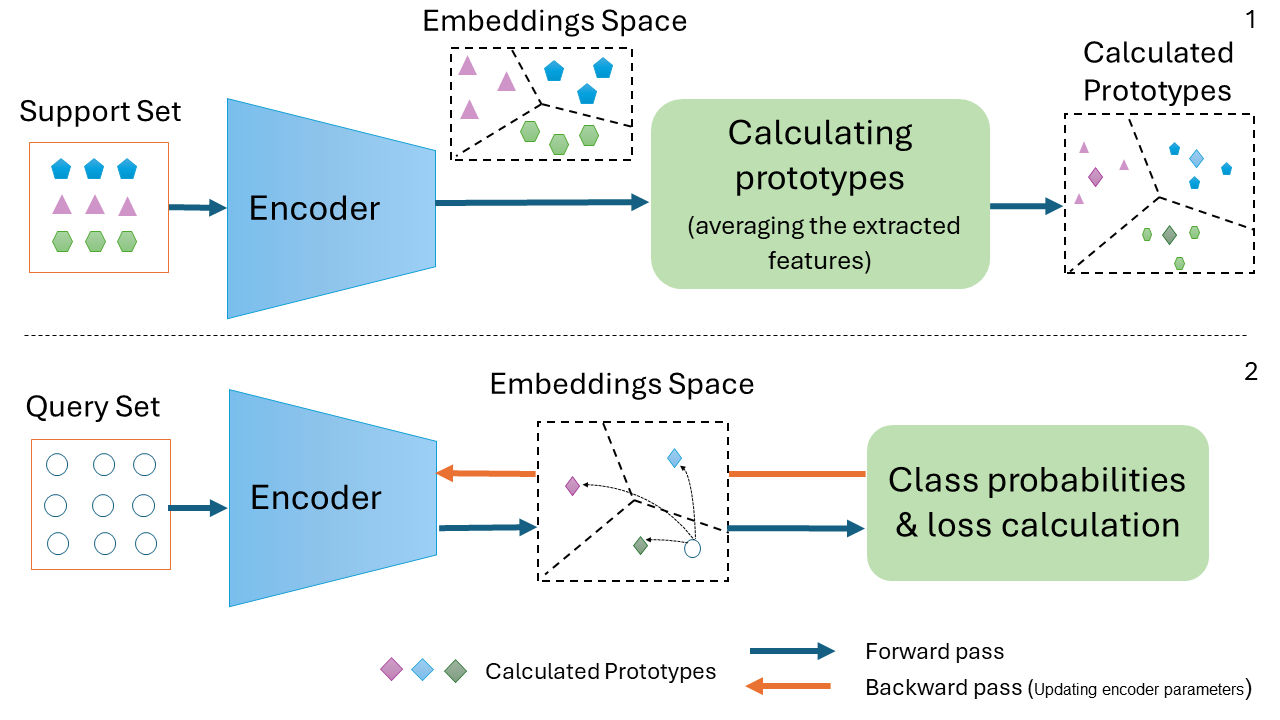}
\caption{Prototypical Network Architecture and Training \cite{protoNet}.}
\label{protoFig}
\end{figure}

Recently, meta-learning approaches such as prototypical networks (PN), have gained significant attention for their effectiveness in data-limited data and domain adaptation scenarios. PN \cite{protoNet} excel at generalizing across classes with minimal labeled data, making them useful in environments where acquiring comprehensive labeled datasets is challenging. PN work by learning how to create an embedded representation (\textit{prototypes}) for each class based on a small set of labeled examples, a paradigm that is referred to as \textit{few-shot} learning. 

Based on these prototypes, it classifies a new sample by finding the closest prototype to this sample's embedded feature. As shown in Figure \ref{protoFig}, during training, the network learns an embedding space that groups samples of the same class closer together while separating those from different classes. Given that PNs are well-regarded for few-shot learning, we explore their potential for generalizing to unseen classes in the AoA estimation task. This investigation aims to answer two main questions: how effectively can a PN learn distinctive feature representations for AoA classification, and to what extent can this learned feature space generalize to novel angles not encountered during training? We hypothesize that the prototypical network’s ability to generate discriminative embeddings will support accurate AoA classification, even under few-shot conditions of unseen angles.

The main contributions of this paper are:
\begin{itemize}
    \item We propose \textbf{ProtoAoA}, a few-shot learning approach that uses real-world IQ signals to efficiently perform AoA estimation with a very limited training dataset. To the best of our knowledge, this study is the first to adopt prototypical networks for efficient AoA classification. The proposed approach is extensible to other tasks offering a novel perspective on addressing data scarcity in AI for wireless communications.
    \item Our work demonstrates the robustness and effectiveness of prototypical networks across various data-limited conditions. We evaluate the network's performance in progressively challenging data setups, where it encounters an increasing number of unseen classes, evaluating the trade-off between data scarcity and accuracy.  Our findings show that training on only 23\% of the AoA dataset classes can attain a mean absolute error (MAE) of 3$^{\circ}$ with only 4-shots of training on the unseen angles and an MAE of 2$^{\circ}$ with 32-shots of training data.
    \item We provide the aforementioned analysis with a real-world dataset collected on a software defined radio (SDR) testbed under diverse sample rates and modulations.

\end{itemize}

The rest of this paper is organized as follows. Section \Romannum{2} summarizes of related work. In Section \Romannum{3}, we describe our testbed, dataset, and proposed ProtoAoA methodology. Section \Romannum{4} discusses our results and findings, and we conclude the paper in Section \Romannum{5}.

\section{Background \& Related Work}

\textbf{Angle-of-Arrival Estimation.} 
Angle-of-Arrival (AoA) estimation is the task of deducing the direction from which a signal arrives at a receiver. It involves determining the angle between the source (e.g., user equipment) and the receiver (e.g., base station). AoA estimation is critical for accurately pinpointing the source's direction and is often a step in positioning and localization or in refining the beam's focus \cite{apps2, AoA-BeamSelection}. The MUSIC algorithm \cite{MUSIC} is commonly used to estimate the AoA by analyzing the phase differences in channel state information (CSI) across multiple antennas. However, MUSIC faces several challenges, particularly under multipath conditions, where the assumption that the signal and noise subspaces are mutually orthogonal may be violated. In this context, methods such as AoA-net \cite{AoA-net} and DeepAoANet \cite{DeepAoANet} have been proposed for AoA estimation.

In particular, DeepAoANet \cite{DeepAoANet}, a data-driven method, was proposed to estimate the number of signals received by an antenna array and their respective AoAs. Its two developed DL architectures outperform the standard MUSIC algorithm. However, the approach requires computing the antennas' covariance matrix as input to the DL model. Similarly, in another study\cite{ML-AoA}, ML techniques were applied to efficiently estimate AoA without introducing significant additional computational overhead, achieving at least 20\% improvement over the baseline MUSIC algorithm. However, building such models typically requires attaining a comprehensive data set with respect to the possible classes of angles, which is a challenge often faced in real-world scenarios.

\begin{figure}[!t]
  \centering
  \includegraphics[width=0.41\textwidth]{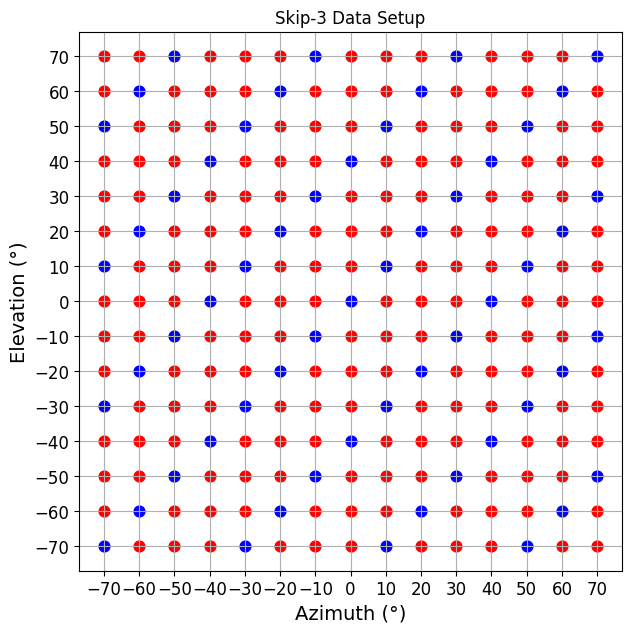}
  \caption{Proposed "Skip-3" few shot learning AoA data set-up: blue dots represent angles seen during training, while red ones represent previously unseen angles.}
  \label{fig_skip3}
\end{figure}
\section{ProtoAoA: Few-Shot Learning of Angle-of-Arrival Estimation}
\textbf{Prototypical Networks.}
Recently, meta-learning methods have acquired significant attention as a means to develop adaptable solutions for wireless communication challenges, particularly facing the limited-data challenge. In \cite{proto-signal-recongition}, the authors highlight the effectiveness of the prototypical network for few-shot learning in signal recognition tasks with high accuracy. In another work \cite{proto-AMC}, prototypical network was adopted with residual attention to achieve effective and robust performance in the modulation classification problem. More recently, their use has expanded to a broader range of wireless applications, including fine-grained few-shot signal modulation classification \cite{FTPNet}, wireless beam prediction under unseen antenna configurations \cite{ProtoBeam}, CSI-based adaptive beam prediction \cite{HybridBeamProto}, and lightweight, generalizable few-shot AoA estimation \cite{Lightweight_AoA}. Despite these promising results, the use of prototypical networks for sparse-data AoA estimation directly from raw IQ measurements remains limited, which is the focus of this paper.

\subsection{SDR Testbed and Dataset}

\begin{figure}[!t]
  \centering
  \includegraphics[scale=0.64]{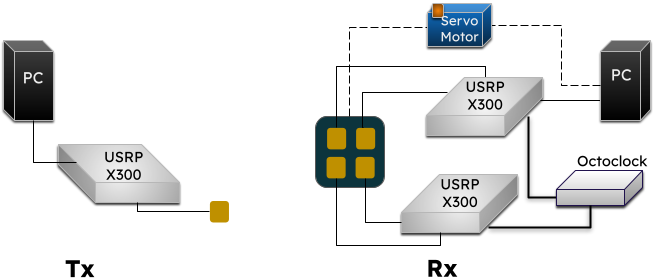}
  \caption{A diagram of the testbed used to create the dataset used through our study.}
  \label{fig:testbed}
\end{figure}

The dataset used in this study was generated using a software defined radio (SDR) testbed. It contains IQ wireless signal data captured outdoors using a USRP X300 transmitter and two synchronized USRP X300 receivers using an Ettus Octoclock CDA-2990. Each of the X300 USRPs have 2 Tx and 2 Rx channels and they were equipped with 5.88 GHz patch antennas. There is a total of 4 Rx receiver antennas and the IQ signals at different AoA were captured across the 4 channels. A diagram of the testbed is shown in Figure \ref{fig:testbed}

To cover various angles of arrival, servo motors adjusted the azimuth and elevation angles of the receiver antennas, spanning angles from -70° to 70° in 10° increments. The dataset also includes signals with different modulation schemes (like 16-QAM, 64-QAM, BPSK) and various sampling rates to add diversity to the collected data.

The dataset is comprised of 225 unique angle combinations, which we refer to as "\textbf{angles}". These are shown in Figure \ref{fig_skip3} (15 x 15 angle combinations, both the red and blue angle combinations were collected by the testbed). The difference between the red and blue angles will be described next.

\subsection{Problem Statement: Few-shot Angle-of-Arrival Estimation}

The primary objective is to classify a stream of received IQ samples from the 4 received channels to one of the 255 angles in Figure \ref{fig_skip3}. The \textbf{few-shot learning} objective, that is the focus of this paper, is to accomplish this with very limited data collection and training. 
Such a scenario is illustrated in \ref{fig_skip3} where the blue angle set constitutes angle combinations where received IQ data streams were collected and used for deep learning training. This translates to having to capture IQ data from only 57 angle locations out of the 255 angle combinations. Only a few shots of data are captured at the remaining red angle combinations. We refer to this set-up as the "\textbf{Skip-3}" few-shot learning setup since the 3 surrounding angles (in x and y) of each blue angle is "skipped" during training. In a similar vein, "\textbf{Skip-2}" and "\textbf{Skip-1}" few-shot learning setups can be evaluated. Next, we discuss how we propose to solve this problem with prototypical networks.

\subsection{Prototypical Networks for Sparse-Data AoA Estimation}

Prototypical networks aim to learn an embedding space where data points belonging to the same class are grouped around a central prototype as illustrated in Figure \ref{protoFig}. By utilizing a deep learning encoder, the network transforms raw inputs into this embedding space, effectively mapping each input to a feature representation. Classification is performed by measuring the distance between a query point and these prototypes, with the query point assigned to the class of the nearest prototype.

\begin{figure}[!t]
  \centering
  \includegraphics[scale=0.7,angle=90]{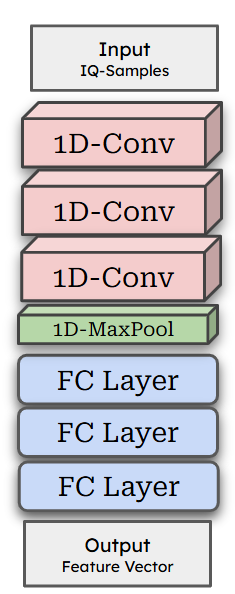}
  \caption{The prototypical network encoder architecture.}
  \label{fig:model}
\end{figure}

\begin{algorithm}[!b]
\caption{Proposed ProtoAoA Training Algorithm}
\label{alg:train}

\textbf{Parameters:}
\begin{itemize}
    \item $N$: number of classes per episode 
    \item $k$: number of support and query samples per class
    \item $N_{ep}$: number of training epochs
\end{itemize}
\textbf{Inputs:}
\begin{itemize}
    \item Training set $\mathcal{D} = \{(x_1, y_1), \ldots, (x_n, y_n)\}$, where $x_i$ is an IQ sample, $y_i \in Y_u$ is the target angle class, $Y_u$ denotes unique classes.
\end{itemize}
\textbf{Outputs: } $\phi$ \Comment{model parameters}

\begin{algorithmic}
\Procedure{Train-ProtoAoA}{$N_{ep}$}
\For{$epoch = 1 \to N_{ep}$}
    \For {$episode = 1 \to len(Y_u)/N$}
        \State $C_{ep} \gets \text{sample } N \text{ classes from } Y_u$ 
        \State TrainEpisode($C_{ep}$)
    \EndFor
\EndFor
\EndProcedure

\Procedure{TrainEpisode}{$C_{ep}, \mathcal{D}$}
\State $S, Q \gets \text{get 2 sets of } k \text{ samples from each class in } C_{ep}$
\State $F_S \gets f_\phi(S)$ \Comment{Forward pass on support set}
\State $F_Q \gets f_\phi(Q)$ \Comment{Forward pass on query set}
\For{$y \in C_{ep}$}
    \State $P_y \gets \frac{1}{k} \sum_{i=1}^{k} F_S(x_{i})$ \Comment{Prototype of $y$}
\EndFor

\State $\hat{y}_Q \gets \operatorname*{arg\,min}_{y'} (softmax( distance(F_Q, P_{y'})))$
\State $L \gets \frac{1}{N \times k} \sum crossEntropy(y_Q, \hat{y}_Q)$

\State $\phi \gets \phi - \alpha \cdot \nabla_\phi L$ \Comment{Update model parameters}
\EndProcedure

\end{algorithmic}
\end{algorithm}

In our approach, \textbf{ProtoAoA}, we employ this framework to classify AoA using IQ input data streams. Each unique \textbf{angle} (azimuth, elevation) is treated as a distinct class. During training, prototypes are generated for specific angles, representing a diverse set of signal characteristics. The intuition of ProtoAoA is that the embeddings generated by the PN are expected to enable a high classification performance when tested on previously unseen angles with only a few-shots. It is important to note that \textbf{no retraining is needed} in ProtoAoA, only new prototypes are created for the unseen angles with a few-shots. 

Following the typical training procedure for prototypical networks \cite{protoNet}, we use the $k$-way classification method where $k$ denotes the number of classes selected for each training episode. Within each selected class, we randomly pick $N$ samples per class for training the network encoder. The encoder and training details are discussed in the following subsections. The complete training process is detailed in Algorithm \ref{alg:train}.

\subsubsection{\textbf{Network Encoder Architecture}}

The network encoder in our architecture is a simple multi-layer convolutional model designed to process complex IQ sample data. The model takes as input a signal array of shape (4, $n$), representing the four-channel IQ samples where $n$ is the number of concatenated IQ samples. The encoder begins with a series of three 1D convolutional layers, each progressively expanding the feature space from the initial input channels to eight times that size. These convolutional layers apply learnable filters, effectively capturing distinctive features relevant to AoA classification. 

Following the convolutional layers, the model applies a max-pooling layer, which reduces the features' dimensions while preserving key features. The output is then flattened into a 1D array, ready for three successive fully connected layers that reduce the representation to the desired output dimensionality which is set to 10. The encoder architecture is presented in Figure \ref{fig:model}.

During training, $N$ samples from $k$ classes are randomly selected to form the \textit{support} set, with an additional $N$ samples from the same $k$ classes forming the \textit{query} set. Both, \textit{support} set and \textit{query} set are passed through the network encoder to extract features via a traditional forward pass. For the \textit{support} samples only, each sample’s class label is then used to group the class's features and then create class prototypes based on these extracted features form the \textit{support} set. 

\subsubsection{\textbf{Prototype Calculation}}

Once \textit{support} features are extracted, class prototypes are computed by averaging the embeddings of all features corresponding to each class, as defined in Equation \ref{eq:prototypes}. Through this step, we represent each class by a single prototype, reducing the IQ samples into $k$ prototypes.

\begin{equation}
P_{y} = \frac{1}{|S_{y}|} \sum_{(x_i, y) \in S_{y}} f_\phi (x_i)
\label{eq:prototypes}
\end{equation}

where $S_{y}$ denotes the set of IQ samples for the target angle $y$, $ x_i \in \mathbb{R}^D $ is an input IQ sample vector with dimension $D$ (($4$, $n$) in our dataset), and $f_\phi$ is the network encoder function that maps the IQ samples $x_i$ to the embedding space. The resulting $P_{y}$ is then the prototype for the target class $y$.

\begin{algorithm}[!b]
\caption{ProtoAoA K-Shot Learning \& Testing Algorithm}
\label{alg:test}

\textbf{Inputs:}
\begin{itemize}
    \item Trained model $f_\phi$
    \item Test set $\mathcal{T} = \{(x_1, y_1), \ldots, (x_n, y_n)\}$, where $x_i$ is an IQ sample, $y_i$ is the target label, $Y_{u}$ denotes unique labels.
    \item $\mathcal{K}$: number of shots (support examples) per class
\end{itemize}

\textbf{Outputs:} $\hat{y}_Q$ \Comment{Predicted Classes}

\begin{algorithmic}

\Procedure{Test-ProtoAoA}{$x$}
    \For{$t = 0 \to len(T)/\mathcal{K}$}
        \State $\mathcal{T}_{s} \gets \text{Randomly sample } k \text{ samples from each class}$
       
        \For{$y \in Y_u$}
            \State $F_S \gets f_\phi(\mathcal{T}_{s}(y))$ \Comment{Forward pass}
            \State $P_y \gets \frac{1}{k} \sum_{i=1}^{\mathcal{K}} F_{S}(i)$ \Comment{Prototype for class $y$}
        \EndFor

        \State $\mathcal{T}_{q} \gets \mathcal{T} \setminus \mathcal{T}_{s}$

        \State $(X_Q, y_Q) \gets T_{q}$
        \State $F_Q \gets f_\phi(X_Q)$ \Comment{Forward pass}
        \State $\hat{y}_Q \gets \operatorname*{arg\,min}_{y'} (softmax( distance(F_Q, P_{y'})))$
    \EndFor
\EndProcedure

\end{algorithmic}
\end{algorithm}

\subsubsection{\textbf{Classification and Encoder Update}}

After forming the prototypes, the extracted \textit{query} features are classified into one of the known-for-training classes based on their proximity to the prototypes. First, the distances (e.g., Euclidean distances) from the query's feature to each class prototypes is calculated. Then, a \textit{Softmax} function is applied over the negative values of those distances to normalize them and convert them into probabilities as given in Equation \ref{eq:probabilities}. 

\begin{equation}
p_\phi(\hat{y} = y|x_{i}) = \frac{\exp(-d(f_\phi(x_{i}), P_{y}))}{\sum_{y'} \exp(-d(f_\phi(x), P_{y'}))}
\label{eq:probabilities}
\end{equation}

where $d$ is the chosen distance metric, and the sum in the denominator extends over all classes. 

The model predicts the class of a query IQ sample $x_{i}$ based on the highest probability from these calculations. For each query sample, we compute the classification loss using the \textit{cross-entropy} function, which compares the predicted probabilities with the true class label. This drives the model’s updates during backpropagation.

The average loss over all query samples is the used to adjust model parameters through traditional gradient descent, with the goal of minimizing the loss and enhancing classification accuracy. After a number of epochs, the network encoder generates representative embeddings for each angle combination that capture the defining features to maps IQ data to corresponding AoA.

\begin{figure*}[!tb]
  \centering
  \subfigure[Skip-1: \textit{Accuracy}]{\includegraphics[width=0.325\textwidth]{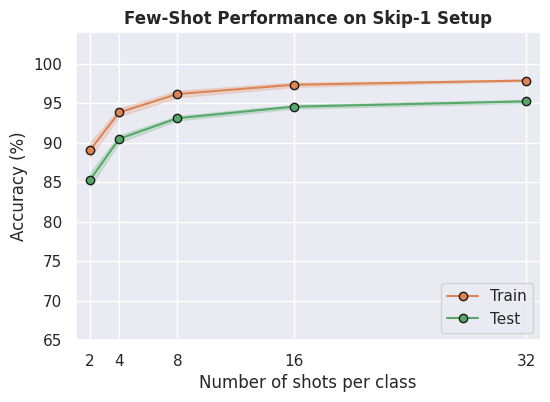}\label{fig:skip1_acc}}
  \subfigure[Skip-2: \textit{Accuracy}]{\includegraphics[width=0.325\textwidth]{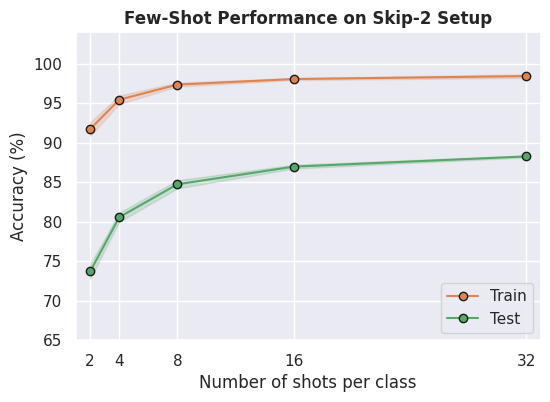}\label{fig:skip2_acc}}
  \subfigure[Skip-3: \textit{Accuracy}]{\includegraphics[width=0.325\textwidth]{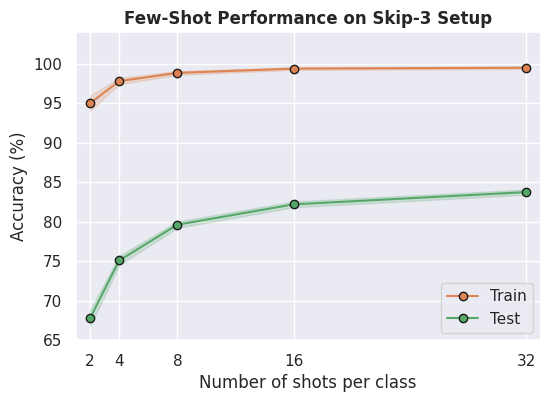}\label{fig:skip3_acc}}
  \subfigure[Skip-1: \textit{MAE}]{\includegraphics[width=0.325\textwidth]{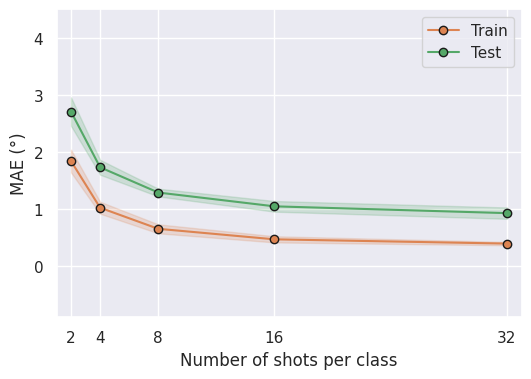}\label{fig:skip1_mae}}
  \subfigure[Skip-2: \textit{MAE}]{\includegraphics[width=0.325\textwidth]{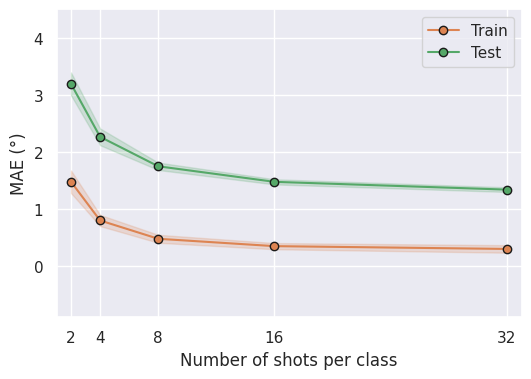}\label{fig:skip2_mae}}
  \subfigure[Skip-3: \textit{MAE}]{\includegraphics[width=0.325\textwidth]{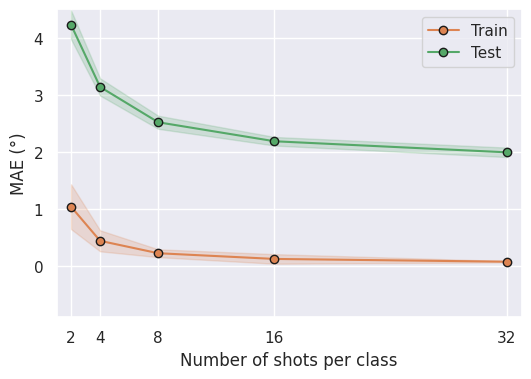}\label{fig:skip3_mae}}
  
  \caption{Accuracy and MAE performance on few-shot testing with training and test data across different data setups.}
  \label{fig:PN_train_test}
\end{figure*}
\subsubsection{\textbf{Few-Shot Learning and Testing Engine}}

After training, the network encoder is used to embed samples from any class into the feature space learned by the encoder. To evaluate the model's adaptability to new, previously unseen angles, we conduct few-shot testing as detailed in Algorithm \ref{alg:test}.

In this procedure, $\mathcal{K}$ samples are randomly selected without replacement from each class to form the \textit{support} set, while the remaining samples form the \textit{query} set. Then, the \textit{support} samples are used to generate the prototypes similar to the prototypes calculation during the training procedure. Once these prototypes are obtained, the same nearest-neighbor method is applied to assign a class to each \textit{query} sample.

To ensure the robustness of the model's performance, the testing process is repeated multiple times, with different samples chosen for the \textit{support} set each iteration. In our method, every sample is guaranteed to be picked in the \textit{support} set once, providing comprehensive coverage across all samples and robust performance assessment.


\section{Results and Discussion}
In this section, we present the results of our experiment of prototypical network training and following it by few-shot testing. Throughout the study, we use the classification accuracy and mean absolute error (MAE) as evaluation metrics. The classification accuracy is the percentage of correctly classified angles from the 255 possible angles, and the MAE captures the average prediction error in degrees. 
The results are obtained on all the \textbf{Skip-1}, \textbf{Skip-2} and \textbf{Skip-3} scenarios discussed in Section III.B.

\subsection{Experimental Setup}
For training and testing, we used a system with an Intel Xeon CPU featuring 32 cores (64 threads) and a base clock speed of 2.30 GHz, supported by 40 GiB of RAM and an NVIDIA Tesla T4 GPU. The primary framework used for code development was PyTorch.

\subsection{ProtoAoA Performance}
\textbf{Classification Accuracy.} Figure \ref{fig:PN_train_test} presents the classification accuracies, and mean absolute errors (MAE) for training and unseen test data. 
Across all data setups, the network achieves at least 90\% accuracy on the classes observed during training. The classification accuracy increases logarithmically with the number of samples used as the \textit{support} set to generate the prototypes ($\mathcal{K}$). As shown in Figures \ref{fig:skip1_acc},\ref{fig:skip2_acc} and \ref{fig:skip3_acc}, the accuracy can reach up to 99\% with $\mathcal{K} \in \{16, 32\}$.

Meanwhile, on previously unseen classes, accuracy generally declines compared to the performance on seen ones during training. The accuracy green curves in Figures \ref{fig:skip1_acc},\ref{fig:skip2_acc} and \ref{fig:skip3_acc} indicate that the test data performance has 10-20\% reduction relative to the training data. This reduction, although it is relatively small, reflects the information gap between the features learned on training and test classes. This reduction gap, however, widens as we move from Skip-1 to Skip-3 setups. Nonetheless, the accuracy can still reach up to 90\% in Skip-2 setup and around 95\% with Skip-1 setup. It is worth noting that a 90\% AoA accuracy can still results in an overall low MAE as we discuss next. 

\textbf{Mean Absolute Error (MAE).} On training data, the MAE values follow a similar overall trend to the classification accuracy. The MAE is initially low with only $\mathcal{K} = 2$, and decreases as $\mathcal{K}$ increases. For example, the orange MAE curve in Figure \ref{fig:skip1_mae} starts with a mean value of $2^{\circ}$, with a notably small standard deviation. The MAE continues to decrease, reaching as low as $0.1^{\circ}$ in some cases, as seen in Figure \ref{fig:skip3_mae}.

On the other hand, the MAE values reported for the previously unseen testing classes are slightly higher than those for the training data. Similar to the classification accuracy, this increase is expected due to the knowledge gap between training and testing classes. The highest MAE observed is $4^{\circ}$ in the Skip-3 setup with $\mathcal{K} = 2$. However, the MAE decreases to approximately $2^{\circ}$, $1.5^{\circ}$, and $1^{\circ}$ in the Skip-3, Skip-2, and Skip-1 setups, respectively.

\begin{figure}
  \centering

  \subfigure{\includegraphics[width=0.45\textwidth]{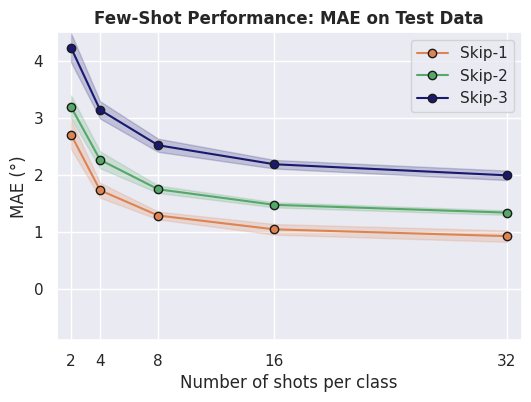}\label{fig:setups_mae_test}}

  \caption{Comparison of MAE performance on few-shot testing with unseen test data across different data setups.}
  \label{fig:compare_test}
\end{figure}

\textbf{Data Efficiency Trade-offs.} To further evaluate the prototypical network's performance on previously unseen classes, we compare its MAE performance on test data across the three distinct data setups, as shown in Figure \ref{fig:compare_test}. From the MAE curves, we observe that the performance on the Skip-1 data setup outperforms that of the two other setups. With only four samples to generate the prototypes, the model achieve an MAE of less than $2^{\circ}$ which corresponds to 90\% classification accuracy as well. Increasing the number of samples to 32 ($\mathcal{K} = 32$) further reduces the MAE value to nearly $1^{\circ}$.

Comparing Skip-2 to Skip-3 results also highlights superior performance on Skip-2. In the latter setup, the prototypical network achieves around 85\% classification accuracy with $\mathcal{K} = 8$, corresponding to an MAE of less than $2^{\circ}$. Meanwhile, reaching the same performance on the Skip-3 setup requires about four times as many samples per class ($\mathcal{K} = 32$).

This gradual decline in performance from Skip-1 to Skip-3 is attributed to the increasing number of classes unseen during training. Such behavior is common in machine learning tasks, where more unseen classes lead to greater generalization challenges. 

\begin{figure}
  \centering
  \includegraphics[width=0.45\textwidth]{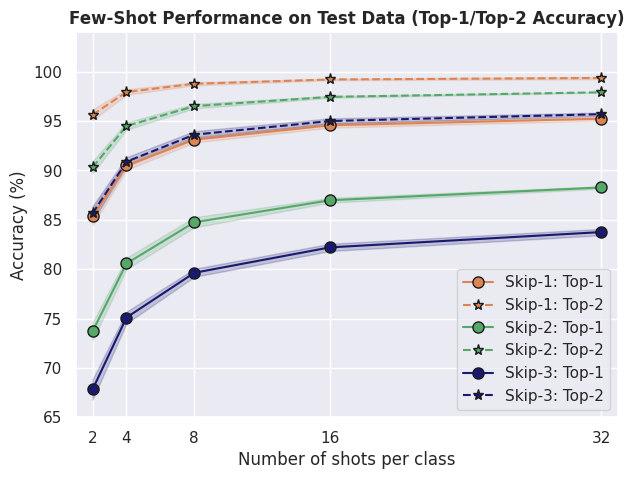}
  \caption{Comparison of Top-1 and Top-2 classification accuracies performance on few-shot testing with unseen test data across different data setups.}
  \label{fig:top2}
\end{figure}

\textbf{Top-2 Accuracy.} In Figure \ref{fig:top2}, we analyzed the top-1 and top-2 accuracy curves across different data setups. Expectedly, the top-2 accuracy consistently exceeds the top-1 accuracy, achieving over 85\% in all setups when only two labeled samples are provided. This improvement is especially notable in the challenging Skip-3 setup, where we notice more than 15\% increase of accuracy compared to top-1 case. This result suggests that the network's predictions are often very close to the correct angle, even when misclassified under top-1 criteria. This trend indicates that the network's representation may struggle slightly with distinguishing neighboring angles, a challenge that can be further addressed in future research.

To sum up, the prototypical network demonstrates strong adaptability and effective performance across different data setups, requiring remarkably few samples to achieve high classification accuracy and low MAE. These findings emphasize the prototypical network's potential in real-world, data-constrained applications, showcasing its ability to generalize effectively and quickly with minimal available data.

\section{Conclusion \& Future Work}
This paper shows the remarkable potential of using meta-learning approaches with few-shot learning for building AI solutions in data-limited wireless environments. Using prototypical networks, we achieved high  accuracy and low mean absolute error in the angle-of-arrival (AoA) classification task with sparse data. Notably, these results were attained with a relatively simple model architecture, showcasing the model's efficiency. This work paves the way for further exploration of prototypical networks in AoA estimation. Future research could focus on enhancing classification performance and refining the model's sensitivity to the relative proximity of similar prototypes. Another promising direction is to investigate whether raw-IQ foundation models, such as IQFM, can further improve label efficiency for sparse-data AoA estimation by providing reusable representations prior to few-shot adaptation \cite{IQFM}.



\bibliographystyle{ieeetr}

\bibliography{IEEEabrv,main}

@ARTICLE{apps2,
  author={Kuznetsov, Yury and Baev, Andrey and Konovalyuk, Maxim and Gorbunova, Anastasia and Russer, Johannes A.},
  journal={IEEE Transactions on Electromagnetic Compatibility}, 
  title={Autocorrelation Analysis and Near-Field Localization of the Radiating Sources With Cyclostationary Properties}, 
  year={2020},
  volume={62},
  number={5},
  pages={2186-2195},
  doi={10.1109/TEMC.2019.2946748}}

@ARTICLE{MUSIC,
  author={Schmidt, R.},
  journal={IEEE Transactions on Antennas and Propagation}, 
  title={Multiple emitter location and signal parameter estimation}, 
  year={1986},
  volume={34},
  number={3},
  pages={276-280},
  doi={10.1109/TAP.1986.1143830}}

@article{AoA-net,
  title={AoA-net: Estimating Angle-of-arrival Using Wi-Fi Channel State Information Based on Deep Neural Networks with Subcarrier Selection},
  author={Kumrai, Teerawat and Cai, Zesheng and Maekawa, Takuya and Hara, Takahiro and Ohara, Kazuya and Murakami, Tomoki and Abeysekera, Hirantha},
  journal={Journal of Information Processing},
  volume={32},
  pages={863--872},
  year={2024},
  publisher={Information Processing Society of Japan}
}

@ARTICLE{DeepAoANet,
  author={Dai, Zhuangzhuang and He, Yuhang and Tran, Vu and Trigoni, Niki and Markham, Andrew},
  journal={IEEE Access}, 
  title={DeepAoANet: Learning Angle of Arrival From Software Defined Radios With Deep Neural Networks}, 
  year={2022},
  volume={10},
  number={},
  pages={3164-3176},
  doi={10.1109/ACCESS.2021.3140146}}

@ARTICLE{AoA-BeamSelection,
  author={Antón-Haro, Carles and Mestre, Xavier},
  journal={IEEE Access}, 
  title={Learning and Data-Driven Beam Selection for mmWave Communications: An Angle of Arrival-Based Approach}, 
  year={2019},
  volume={7},
  number={},
  pages={20404-20415},
  doi={10.1109/ACCESS.2019.2895594}}

@ARTICLE{ML-AoA,
  author={Khan, Aftab and Wang, Stephen and Zhu, Ziming},
  journal={IEEE Communications Letters}, 
  title={Angle-of-Arrival Estimation Using an Adaptive Machine Learning Framework}, 
  year={2019},
  volume={23},
  number={2},
  pages={294-297},
  doi={10.1109/LCOMM.2018.2884464}}

@INPROCEEDINGS{proto-signal-recongition,
  author={Wang, Hanhong and Qi, Lin and Han, Yu and Lin, Yun},
  booktitle={2022 9th International Conference on Dependable Systems and Their Applications (DSA)}, 
  title={Prototypical Network for Few-Shot Signal Recognition}, 
  year={2022},
  volume={},
  number={},
  pages={980-985},
  doi={10.1109/DSA56465.2022.00138}}

@article{proto-AMC,
  title={Prototypical Network with Residual Attention for Modulation Classification of Wireless Communication Signals},
  author={Zang, Bo and Gou, Xiaopeng and Zhu, Zhigang and Long, Lulan and Zhang, Haotian},
  journal={Electronics},
  volume={12},
  number={24},
  pages={5005},
  year={2023},
  publisher={MDPI}
}

@article{AoA-cell,
title = {A scalable fingerprint-based angle-of-arrival machine learning approach for cellular mobile radio localization},
journal = {Computer Communications},
volume = {157},
pages = {92-101},
year = {2020},
issn = {0140-3664},
doi = {https://doi.org/10.1016/j.comcom.2020.04.014},
url = {https://www.sciencedirect.com/science/article/pii/S0140366419314367},
author = {Robson D.A. Timoteo and Daniel C. Cunha},
}

@article{protoNet,
  title={Prototypical networks for few-shot learning},
  author={Snell, Jake and Swersky, Kevin and Zemel, Richard},
  journal={Advances in neural information processing systems},
  volume={30},
  year={2017}
}

@INPROCEEDINGS{Lightweight_AoA,
  author={Mashaal, Omar and Mohammed, Elsayed and Digby, Alec and Leone, Pasquale and Swersky, Lorne and Eshaghbeigi, Ashkan and Abou-Zeid, Hatem},
  booktitle={ICC 2025 - IEEE International Conference on Communications}, 
  title={Lightweight and Generalizable AoA Estimation for IoT: A Novel Few-Shot Learning Approach}, 
  year={2025},
  volume={},
  number={},
  pages={686-691},
  doi={10.1109/ICC52391.2025.11160710}
}

@INPROCEEDINGS{ProtoBeam,
  author={Mashaal, Omar and Mohammed, Elsayed and Digby, Alec and Swersky, Lorne and Eshaghbeigi, Ashkan and Abou-Zeid, Hatem},
  booktitle={GLOBECOM 2024 - 2024 IEEE Global Communications Conference}, 
  title={ProtoBeam: Generalizing Deep Beam Prediction to Unseen Antennas using Prototypical Networks}, 
  year={2024},
  volume={},
  number={},
  pages={133-138},
  doi={10.1109/GLOBECOM52923.2024.10901744}
}

@INPROCEEDINGS{HybridBeamProto,
  author={Thiangkate, Chaiyawut and Nishiyama, Shimpei and Chakraborty, Dipanita and Okada, Minoru and Thonglek, Kundjanasith and Leelaprute, Pattara and Rungsawang, Arnon and Manaskasemsak, Bundit and Chamnongthai, Kosin},
  booktitle={2025 24th International Symposium on Communications and Information Technologies (ISCIT)}, 
  title={Hybrid Feature Fused Few-Shot Learning for CSI-Based Adaptive Beam Prediction}, 
  year={2025},
  volume={},
  number={},
  pages={13-18},
  doi={10.1109/ISCIT67082.2025.11231746}
}

@article{BeamHeterogeneity,
  author={Zeulin, Nikita and Galinina, Olga and Kilinc, Ibrahim and Andreev, Sergey and Heath Jr, Robert W},

  title={Rethinking Beam Management: Generalization Limits Under Hardware Heterogeneity},
  journal={arXiv preprint arXiv:2602.18151},
  year={2026}
}

@ARTICLE{IQFM,
  author={Mashaal, Omar and Abou-Zeid, Hatem},
  journal={IEEE Open Journal of the Communications Society}, 
  title={IQFM—A Wireless Foundation Model for I/Q Streams in AI-Native 6G}, 
  year={2026},
  volume={7},
  number={},
  pages={1426-1441},

  doi={10.1109/OJCOMS.2026.3661435}}

@ARTICLE{FTPNet,
  author={Feng, Shuai and Wang, Yatong and Wen, Zhongyi and Xu, Luyan and Yan, Mu},
  journal={IEEE Transactions on Cognitive Communications and Networking}, 
  title={Fine-Grained Transductive Prototypical Network-Based Few-Shot Signal Modulation Classification Using Coarse Labels}, 
  year={2026},
  volume={12},
  number={},
  pages={2189-2204},
  doi={10.1109/TCCN.2025.3594331}
}

\end{document}